\documentclass[twocolumn,showpacs,preprintnumbers,amsmath,amssymb,superscriptaddress]{revtex4}

\usepackage{graphicx}
\usepackage{dcolumn}
\usepackage{bm}

\begin{document}


\title{Gate-controlled non-volatile graphene-ferroelectric memory}

\author{Yi Zheng}
\affiliation{Department of Physics, 2 Science Drive 3, National University of Singapore, Singapore 117542}
\affiliation{NanoCore, 4 Engineering Drive 3, National University of Singapore, Singapore 117576}
\author{Guang-Xin Ni}
\affiliation{Department of Physics, 2 Science Drive 3, National University of Singapore, Singapore 117542}
\author{Chee-Tat Toh}
\affiliation{Department of Physics, 2 Science Drive 3, National University of Singapore, Singapore 117542}
\author{Ming-Gang Zeng}
\affiliation{Department of Physics, 2 Science Drive 3, National University of Singapore, Singapore 117542}
\affiliation{NanoCore, 4 Engineering Drive 3, National University of Singapore, Singapore 117576}
\author{Shu-Ting Chen}
   \affiliation{Institute of Material Research and Engineering (IMRE),
3 Research Link, Singapore 117602}
\author{Kui Yao}
   \affiliation{Institute of Material Research and Engineering (IMRE),
3 Research Link, Singapore 117602}
\author{Barbaros \"{O}zyilmaz}
  \email{phyob@nus.edu.sg}
     \affiliation{Department of Physics, 2 Science Drive 3, National University of Singapore, Singapore 117542}
     \affiliation{NanoCore, 4 Engineering Drive 3, National University of Singapore, Singapore 117576}

\date{\today}

\begin{abstract}
In this letter, we demonstrate a non-volatile memory device in a graphene FET structure using ferroelectric gating. The binary information, i.e. ``1'' and ``0'', is represented by the high and low resistance states of the graphene working channels and is switched by controlling the polarization of the ferroelectric thin film using gate voltage sweep. A non-volatile resistance change exceeding 200\% is achieved in our graphene-ferroelectric hybrid devices. The experimental observations are explained by the electrostatic doping of graphene by electric dipoles at the ferroelectric/graphene interface.
\end{abstract}

\pacs{Valid PACS appear here}
\maketitle
The discovery of graphene in 2004 \cite{Novoselov04Science,Novoselov05Nature,Kim05Nature} has triggered enormous experimental and theoretical efforts \cite{Novoselov07NatMater,Neto08RevModernPhys}. As a gapless semiconductor, charge carriers in graphene can be tuned continuously from electrons to holes crossing the charge neutral Dirac point using an external electric field. Unlike conventional semiconductors, the doping process does not influence the mobility of charge carriers in graphene, which can exceed $10^{5}$ cm$^{2}$V$^{-1}$s$^{-1}$ at low temperature \cite{Andrei08NatureNanotech,BolotinKim08PRL}. Such doping-independent mobility leads to the field-dependent conductance in graphene. Based on these two properties, many novel graphene-based device applications have been predicted or demonstrated \cite{Novoselov07NatMater02,Williams07Science,Barbaros07PRL,Barbaros07APL,Goldhaber-Gordon07PRL,Savchenko08NanoLett,Lau08APL,
Trauzettel07NatPhy,Tombros07Nature}, including the heavily-explored graphene-based field-effect transistor (GFET) \cite{Berger04JPCB,Berger06Scince,Echtermeyer08IEEE,Dai08PRL,HanKim07PRL,Avouris08NanoLett}.
However, a paradigm shift in the microelectronics industry from Si to graphene also requires graphene-based memory applications. Despite graphene intrinsically having a high resistance state at the Dirac point and a low resistance state when heavily doped, reports on graphene for non-volatile information storage is rarely seen. This is due to the difficulty in maintaining the resistance states in graphene without an external electric field. One chemical modification approach to achieve non-volatile switching in graphene has been recently proposed by Echtermeyer \textit{et al} \cite{Echtermeyer08IEEE}. Although this method can achieve very high on-off ratio, it alters the unique crystalline structure of graphene upon which many of the extraordinary electronic properties and hence most novel device concepts are based \cite{Novoselov07NatMater,Neto08RevModernPhys}.

In this letter, we show non-volatile switching in graphene by using ferroelectric gating without having to break the lattice symmetry. We demonstrate basic writing and reading processes of this novel graphene-ferroelectric memory device structure combining the field-dependent conductance of graphene with the remnant electric field of ferroelectric thin films. A bistable state of high and low resistance value is realized by controlling the electrical doping level in graphene hysteretically, which is caused by a hysteretic switching of the polarization in the ferroelectric thin film.

\begin{figure}
\begin{center}
\includegraphics[width=3.0in]{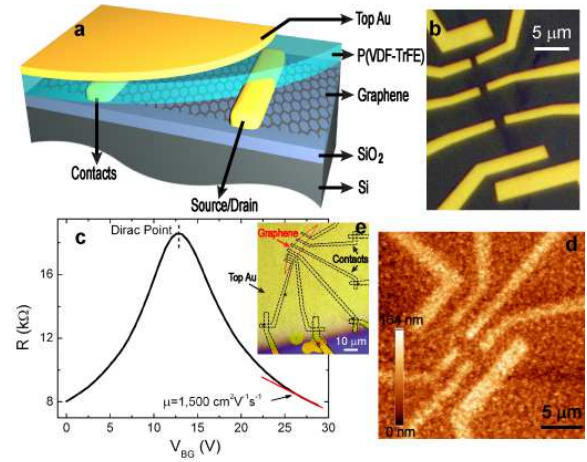}
\end{center}
\caption{(a) Sample geometry of a finished graphene-ferroelectric memory device. (b) Optical image of a graphene sample showing the Hall-bar geometry of the bottom electrodes. (c) R vs $V_\mathrm{BG}$ of the graphene sample before P(VDF-TrFE) coating, measured in two-terminal configuration. (d) AFM image of another graphene sample after P(VDF-TrFE) spin-coating. The contrast comes from the slightly different crystallization of P(VDF-TrFE) on SiO$_2$, Graphene and Au electrodes respectively. (e) Optical image of a finished device.} \label{Fig01}
\end{figure}

The sample geometry of our graphene-ferroelectric memory devices is shown in Fig. \ref{Fig01}a. The bottom electrodes, patterned by electron-beam lithography, were prepared by thermal evaporation of Cr/Au (5/30 nm). A ferroelectric thin film of poly(vinylidene fluoride-trifluoroethylene) (P(VDF-TrFE)) was then spin-coated with a thickness of approximately 0.7 $\mu$m. From atomic force microscopy (AFM), we conclude that P(VDF-TrFE) forms a continuous thin film on graphene devices (Fig. \ref{Fig01}d). For all devices, resistance R vs the bottom gate voltage ($V_\mathrm{BG}$) has been recorded for reference before P(VDF-TrFE) coating. After thermally evaporating the top gate electrodes (Fig. \ref{Fig01}e), samples were electrically characterized at room temperature in vacuum in a four-contact configuration using a lock-in amplifier with an ac excitation current of 10 nA. The number of graphene layers is confirmed by Raman spectroscopy. In total, we have successfully studied 15 samples. For the representative sample used here, the charge carrier mobility before P(VDF-TrFE) is $\sim1500$ cm$^{2}$V$^{-1}$s$^{-1}$, estimated from the linear slope of the R vs $V_\mathrm{BG}$ curve \cite{Novoselov07NatMater02}, as shown in Fig. \ref{Fig01}c.

\begin{figure}
\begin{center}
\includegraphics[width=3in]{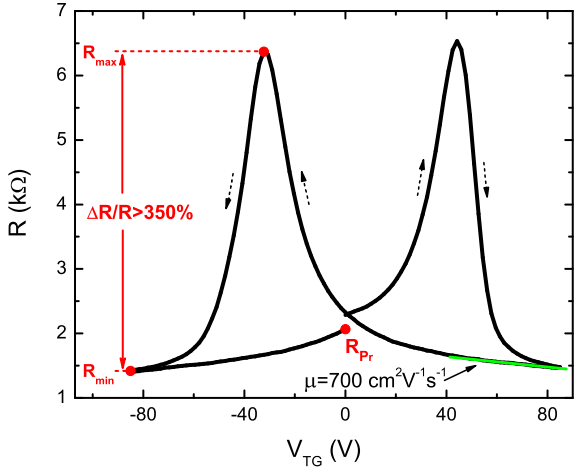}
\end{center}
\caption{Electric hysteresis loop; R as a function of $V_\mathrm{TG}$ for the graphene-ferroelectric sample. The resistance peak at 44 V (-32 V) corresponds to the flipping of electric dipoles in P(VDF-TrFE) from upward (downward) to downward (upward). From the linear part of this curve at high voltage, the charge carrier mobility is estimated to be 700 cm$^{2}$V$^{-1}$s$^{-1}$, taking $\kappa_\mathrm{PVDF}=10$ \cite{Ducharme05IEEE}.} \label{Fig02}
\end{figure}

We now present the main experimental observations. As shown in Fig. \ref{Fig02}, the most important feature for all measured samples is a pronounced hysteresis in resistance measurements when the top gate voltage ($V_\mathrm{TG}$) is swept in a closed loop: 0 V to 85 V, 85 V to -85 V, and finally from -85 V back to 0 V. Similar to the magnetoresistance measurements of a giant magnetoresistance (GMR)/tunneling magnetoresistance (TMR) device, we observe a hysteretic switching between the maximum resistance (R$_\mathrm{max}$) and the minimum resistance (R$_\mathrm{min}$), but now as a function of the applied electric field, E. For the sample in Fig. \ref{Fig02}, the resistance change ratio, $\Delta R/R=(R_\mathrm{max}-R_\mathrm{min})/R_\mathrm{min}$, is larger than 350\%.

\begin{figure}
\begin{center}
\includegraphics[width=3in]{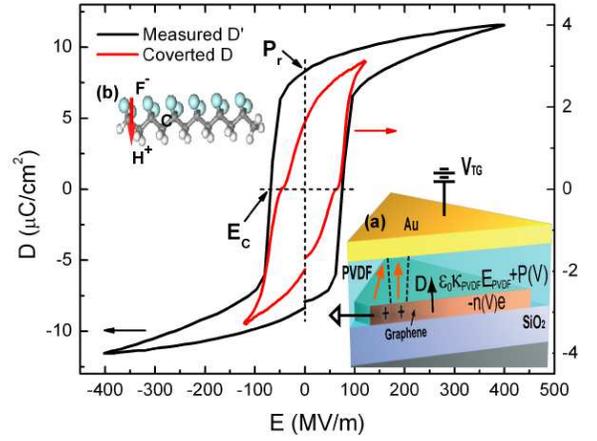}%
\end{center}
\caption{\label{Fig03}D vs $V_{TG}$ characteristics deduced from the R vs $V_{TG}$ curve in Fig. \ref{Fig02}. The black curve represents the experimental measured D' of P(VDF-TrFE) thin film with similar thickness. The kinks near E$_\mathrm{C}$ in D are caused by the assumption of a constant charge carrier mobility, which does not hold near the charge neutral regime. Inset a: the electric displacement continuity equation at ferroelectric/graphene interface. Inset b: a polarized P(VDF-TrFE) molecule. Cyan, grey and white atoms represent fluorine, carbon and hydrogen respectively.}
\end{figure}

This hysteretic behavior and the double peak structure in R vs $V_\mathrm{TG}$ curves are closely related to the polarization of P(VDF-TrFE) thin film. As illustrated in the inset a of Fig. \ref{Fig03}, the continuity of electric displacement field, D, at the ferroelectric/graphene interface requires $\mathrm{D}=\varepsilon_{0} \kappa_\mathrm{ferro} \mathrm{E}_\mathrm{ferro}+\mathrm{P}(V_\mathrm{TG})=-\mathrm{n}(V_\mathrm{TG}) e$, where $\mathrm{E}_\mathrm{ferro}$ and $\mathrm{n}(V_\mathrm{TG})$ are the electric field in ferroelectric and the charge carrier concentration in graphene respectively \cite{Note02}. Here, the dielectric response of the ferroelectric separates into a linear part ($\varepsilon_{0} \kappa_\mathrm{ferro} \mathrm{E}_\mathrm{ferro}$) and a hysteretic part ($\mathrm{P}(V_\mathrm{TG})$). While the linear dielectric part induces electrical doping in graphene with an opposite sign to $V_\mathrm{TG}$, $\mathrm{P}(V_\mathrm{TG})$ can induce electrical doping with either sign. These two components compete with each other such that graphene can remain p-doped (n-doped) even with a positive (negative) $V_\mathrm{TG}$ until either the doping contribution from the linear part exceeds $\mathrm{P}(V_\mathrm{TG})$ or the polarization direction of the ferroelectric is switched. It is this behavior which leads to the hysteretic doping of graphene as a function of $V_\mathrm{TG}$. Considering that the conductance of graphene is $\sigma=\mathrm{n}(V_\mathrm{TG})\mathrm{e}\mu$, the observed resistance hysteresis loop is now directly related to $\mathrm{P}(V_\mathrm{TG})$ by \begin{equation}\label{Equ01}
    \mathrm{D}=\varepsilon_{0} \kappa_\mathrm{ferro} \mathrm{E}_\mathrm{ferro}+\mathrm{P}(V_\mathrm{TG})=-\mathrm{n}(V_\mathrm{TG}) \mathrm{e}=\sigma /\mu.
\end{equation}

Using this equation, we convert the R vs $V_\mathrm{TG}$ curve in Fig. \ref{Fig02} into D vs E characteristics, which is further compared with the direct electric displacement measurement of P(VDF-TrFE) thin film alone, D'. As show in Fig. \ref{Fig03}, D and D' have very similar coercive fields (E$_\mathrm{C}$) of $\sim 50$ MV/m, consistent with the typical E$_\mathrm{C}$ reported in literature \cite{Furukawa06IEEE}. This agreement strongly suggests that the hysteresis observed in the transport measurements is indeed caused by the hysteretic polarization of the ferroelectric gate dielectric. From Fig. \ref{Fig03}, we can also see that the left and right resistance peaks in Fig. \ref{Fig02} correspond to the flipping of electric dipoles from upward to downward and from downward to upward, respectively, while R$_\mathrm{min}$ in Fig. \ref{Fig02} is related to the maximum polarization point in Fig. \ref{Fig03}. Another important parameter in Fig. \ref{Fig02} is the zero-field resistance R$_\mathrm{P_{r}}$, corresponding to a remnant polarization P$_\mathrm{r}$ of P(VDF-TrFE) in Fig. \ref{Fig03}.

\begin{figure}
\begin{center}
\includegraphics[width=3.5in]{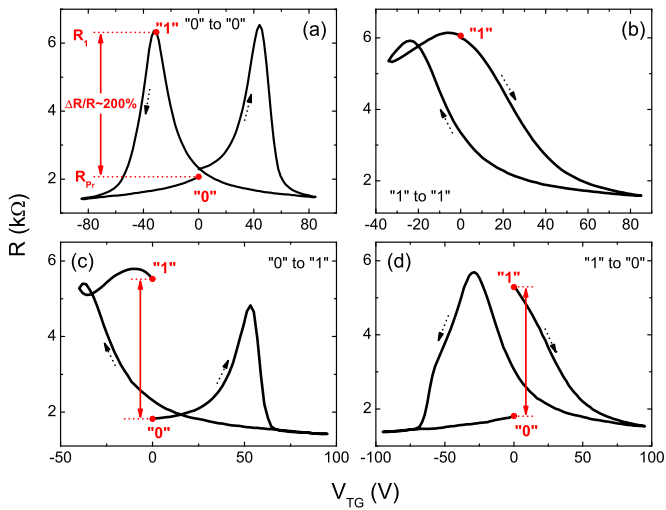}
\end{center}
\caption{(a) Writing ``0'' into graphene-ferroelectric memory by a full loop sweep of $V_\mathrm{TG}$ ($\pm85$ V). The memory bit is in ``0'' state before writing. (b) Writing ``1'' into graphene-ferroelectric memory by an asymmetrical loop sweep of $V_\mathrm{TG}$ (85 V to -34 V). The memory bit is in ``1'' state before writing. (c) Writing ``0'' into graphene-ferroelectric memory by a full loop sweep of $V_\mathrm{TG}$. The memory bit is in ``1'' state before writing. (d) Writing ``1'' into graphene-ferroelectric memory by an asymmetrical loop sweep of $V_\mathrm{TG}$. The memory bit is in ``0'' state before writing.} \label{Fig04}
\end{figure}

For device operation, we utilize the maximum resistance peak ($\mathrm{R}_{1}\simeq \mathrm{R}_\mathrm{max}$) as bit ``1'', while bit ``0'' is represented by R$_\mathrm{P_{r}}$. As shown in Fig. \ref{Fig04}a and \ref{Fig04}d, a major hysteresis loop, corresponding to a full symmetrical $V_\mathrm{TG}$ sweep ($\pm V_\mathrm{max}$), can set the memory to ``0'', independent of the existing state. In Fig. \ref{Fig04}a, ``0'' has been rewritten into ``0'', while in Fig. \ref{Fig04}d, the binary information has been reset from ``1'' to ``0''.

In contrast, writing ``1'' into graphene-ferroelectric memory requires a minor hysteresis loop with an asymmetrical $V_\mathrm{TG}$ sweep to minimize the polarization in P(VDF-TrFE) thin film when $V_\mathrm{TG}$ is back to zero. As shown in Fig. \ref{Fig04}b and \ref{Fig04}c, a minor hysteresis loop with $V_\mathrm{max}=85$ V and $-V_\mathrm{max}'=-34$ V can set the resistance state of graphene channel to near R$_\mathrm{max}$, independent on the initial state of ``1'' (Fig. \ref{Fig04}b) or ``0'' (Fig. \ref{Fig04}c).

Thus, using major and minor hysteresis loops, we can realize non-volatile switching in graphene-ferroelectric memory. Note that the switching voltage can be reduced by one order of magnitude by simply scaling the thickness of P(VDF-TrFE) to the range of 100 nm. The difference between the two resistance states ($\Delta$R/R=(R$_{1}$-R$_\mathrm{P_{r}}$)/R$_\mathrm{P_{r}}$) is decided by the difference between the minimum polarization ($\mathrm{P_{min}}$) and the remnant polarization ($\mathrm{P_{r}}$) in P(VDF-TrFE). For the sample discussed here, the resistance change $\Delta$R/R is $\sim200\%$. The read-out of the binary information of graphene-ferroelectric memory can be simply done by measuring the device resistance using an excitation current as low as 1 nA. Note that this hybrid memory structure can in principle retain the high charge carrier mobility in graphene, which is crucial for ultra-fast device applications. In fact, the reading speed of an ideal device can be as fast as several ten femtoseconds when operating at 1 V with a channel length of 1 $\mu$m and charge carrier mobility of $\mu$=200,000 cm$^{2}$V$^{-1}$s$^{-1}$.

Before concluding, we discuss how to improve the performance of graphene-ferroelectric devices. First, $\Delta\mathrm{R/R}$ can be much improved by removing contaminant residues on the graphene surface (enhancing interfacial coupling) and more importantly, by improving the charge carrier mobility in graphene. For an ideal ferroelectric/graphene interface, $\Delta\mathrm{R/R}$ can be as large as $\sim(\frac{1}{\mathrm{P_{min}} \mu'}-\frac{1}{\mathrm{P_{r}}\mu})/\frac{1}{\mathrm{P_{r}} \mu}$, where $\mu'$ is the mobility at P$_\mathrm{min}$. Another important approach is to increase the remnant polarization by applying larger electric field. A better approach would be the preparation of graphene sheets directly on ferroelectric substrates, which would allow the use of other ferroelectric materials with much higher remnant polarization. Other strategy would be to open a band gap in graphene, either by using bilayer graphene or graphene nanoribbons.

In conclusion, we have demonstrated the working principle and real device operation of a novel hybrid non-volatile memory device using graphene and ferroelectric thin film. Reversible non-volatile switching between the high resistance state and the low resistance state in graphene has been realized by choosing major or minor hysteresis loops. Using the electric displacement continuity equation, we show that the resistance hysteresis loop and the switching between the high and low resistance states are due to the electric dipole induced doping in graphene by the ferroelectric thin film. Currently, the resistance change, $\Delta$R/R, exceeds 200\% and can be further improved by improving the quality of ferroelectric/graphene interface, the charge carrier mobility in graphene, and by increasing the remnant polarization of ferroelectric thin film. These make this new memory structure a promising candidate for the next generation of ultra-fast non-volatile memory.

\begin{acknowledgements}
We particularly acknowledge Douwe J. Monsma for many insight and useful discussions of this work. We thank Natarajan Chandrasekhar and Sow Chorng Haur for their help in our experiments, and Yu Ting and Cong Chunxiao for Ramam characterizations of graphene flakes. This work is supported by the Singapore National Research Foundation under NRF RF Award No. NRF-RF2008-07 and by NUS NanoCore.
\end{acknowledgements}

\newpage

\end{document}